\documentstyle[prl,floats,aps,twocolumn,epsf,graphicx]{revtex}
\begin{document}
\twocolumn[\hsize\textwidth\columnwidth\hsize\csname
@twocolumnfalse\endcsname

\title{Universal Bound on Dynamical Relaxation Times and Black-Hole Quasinormal Ringing}
\author{Shahar Hod}
\address{The Ruppin Academic Center, Emeq Hefer 40250, Israel}
\address{and}
\address{The Hadassah Institute, Jerusalem 91010, Israel}
\date{\today}
\maketitle

\begin{abstract}

\ \ \ From information theory and thermodynamic considerations a universal bound on 
the relaxation time $\tau$ of a perturbed system is inferred, 
$\tau \geq \hbar/\pi T$, where $T$ is the system's temperature. 
We prove that black holes comply with the bound; in fact they actually {\it saturate} it. 
Thus, when judged by their relaxation properties, black holes are the most extreme objects in 
nature, having the maximum relaxation rate which is allowed by quantum theory.
\end{abstract}
\bigskip

]

A fundamental problem in thermodynamic and statistical physics is to study the response of a 
system in thermal equilibrium to an outside perturbation \cite{Kubo,Fet,DaSa}. 
In particular, one is typically interested in calculating the 
relaxation timescale at which the perturbed system returns to a stationary, equilibrium configuration. 
Can this relaxation time be made arbitrarily small? That the answer may be negative is 
suggested by the third-law of thermodynamics, according to which 
the relaxation time of a perturbed system is expected to go to infinity in the limit of 
absolute zero of temperature. 
Finite temperature systems are expected to have faster dynamics and shorter 
relaxation times-- how small can these be made? 
In this Letter we use general results from quantum information theory in order to derive a 
fundamental bound on the maximal rate at which a perturbed system approaches thermal equilibrium.

On another front, deep connections between the world of black-hole physics and the realms of thermodynamics and 
information theory were revealed by Hawking's theoretical discovery of black-hole radiation \cite{Haw}, 
and its corresponding temperature and black-hole entropy \cite{Beken1}. These discoveries imply that 
black holes behave as thermodynamic systems in many respects. 
Furthermore, black holes have been proven to be very useful in deriving fundamental, 
{\it static} bounds on information storage \cite{Bekb1,Hooft,Hode1,Bekb2,Hode2,Bekcon}. 
Can one use black holes to obtain deep insights into natural limitations on {\it dynamical} 
relaxation times? 
Indeed one can, as we shall show below.

{\it Quantum information theory and Thermodynamics.---} A fundamental question in quantum information 
theory is what is the maximum rate, $\dot I_{max}$, at which information may 
be transmitted by a signal of duration $\tau$ and energy $\Delta {\cal E}$. 
The answer to this question was already found in the 1960's (see e.g. \cite{Bekinf,Bre}):

\begin{equation}\label{Eq1}
\dot I_{max} = \pi \Delta{\cal E}/\hbar \ln2\  .
\end{equation}
(We use units in which $k_B=G=c=1$.) 

An outside perturbation to a thermodynamic system is 
characterized by energy and entropy changes in the system, $\Delta{\cal E}$ and $\Delta S$, respectively. 
By the complementary relation between entropy and information (entropy as a measure of one's uncertainty or 
lack of information about the actual internal state of the system \cite{Shanon,Gold,Bri,Bekinf}), 
the relation Eq. (\ref{Eq1}) sets an upper bound on the rate of entropy change \cite{Noteln2}, 

\begin{equation}\label{Eq2}
{{\Delta S} \over \tau} \leq \pi \Delta{\cal E} / \hbar \  , 
\end{equation}
where $\tau$ is the characteristic timescale for this dynamical process 
(the relaxation time required for the perturbed system to return to a quiescent state). 

Taking cognizance of the second-law of thermodynamics, one obtains from Eq. (\ref{Eq2})

\begin{equation}\label{Eq3}
\tau_{min} = \hbar/ \pi T\  ,
\end{equation}
where $T$ is the system's temperature. 
Thus, according to quantum theory, a thermodynamic system has (at least) one perturbation mode 
whose relaxation time is $(\pi T/\hbar)^{-1}$, or larger. 
This mode dominates the late-time relaxation dynamics of the system, and determines the 
characteristic timescale for generic perturbations to decay \cite{Note1}.

Typically for laboratory sized systems $\tau T$ is many orders of magnitude larger than $\hbar$. 
For example, the relaxation timescale of a gas composed of particles of mass $m$ 
is of the order of $R/c_s$, where $R$ is the characteristic dimension of the system, and 
$c_s \sim (T/m)^{1/2}$ is the sound velocity. 
Thus, $\tau T \sim R(Tm)^{1/2} \sim 10^{11}\hbar \gg \hbar$ for room temperatures and $R \sim 1$m. 
One therefore wonders whether there are thermodynamic systems in nature 
whose relaxation times are of the same order of magnitude as the minimal relaxation time, $\tau_{min}$, 
allowed by quantum theory? 

The following argument will guide our search for physical systems that come close to challenge 
the dynamical bound, Eq. (\ref{Eq3}): The characteristic 
relaxation time of a thermodynamic system is bounded from below by the 
time it takes for the perturbation to propagate along the system. 
Thus, the relaxation time is bounded by the characteristic size, $R$, of the system. 
This fact motivates a shift of interest to systems whose temperature is of the same order of 
magnitude (or smaller) as the reciprocal of the characteristic size of the system, $\hbar /R$.

The thermodynamic temperature of a Kerr black hole 
is given by the Bekenstein-Hawking temperature \cite{Haw,Beken1}, 
$T_{BH}={{\hbar (r_+ -r_-)} \over {4\pi(r^2_+ +a^2)}}$, where 
$r_{\pm}=M+(M^2-a^2)^{1/2}$ are the black-hole outer and inner horizons, and $M$ and $a$ are the 
black-hole mass and angular momentum per unit mass, respectively. 
One therefore finds that a spherically symmetric Schwarzschild black hole (with $a=0$) 
satisfies the relation $T_{BH} \sim \hbar/r_+$, and may therefore come close to saturate the 
relaxation bound, Eq. (\ref{Eq3}). 
Moreover, it seems that rotating Kerr black holes may even break the relaxation bound, since their 
temperatures are characterized by $T_{BH} \ll \hbar/r_+$ in the extremal limit $r_- \to r_+$ 
($T_{BH} \to 0$). 
We shall now show that black holes conform to the relaxation bound; in fact they 
actually saturate it.

{\it Black-hole relaxation.---} 
The statement that black holes have no hair was introduced by Wheeler \cite {Wheeler} in the early 1970`s. 
The various no-hair theorems state that the external field of a dynamically formed 
black hole (or a perturbed black hole) relaxes to a Kerr-Newman spacetime, 
characterized solely by three parameters: the black-hole mass, charge, and angular momentum. 
This implies that perturbation fields left outside the black hole would either be radiated away to 
infinity, or be swallowed by the black hole. 

This relaxation phase in the dynamics of perturbed black holes is 
characterized by `quasinormal ringing', damped oscillations with a discrete 
spectrum (see e.g. \cite{Nollert1} for a detailed review). 
At late times, all perturbations are
radiated away in a manner reminiscent of the last pure dying tones of
a ringing bell \cite{Press,Cruz,Vish,Davis}. 
Being the characteristic `sound' of the black hole itself, these free oscillations 
are of great importance from the astrophysical point of view. 
They allow a direct way of identifying the spacetime 
parameters (especially, the mass and angular momentum of the black hole). 
This has motivated a flurry of research during the last four decades aiming to compute 
the quasinormal mode (QNM) spectrum of various types of black-hole spacetimes \cite{Nollert1}. 

In addition, the highly damped black-hole resonances \cite{Hod1} have been the subject of much 
recent attention, with the hope that these classical frequencies
may shed some light on the elusive theory of quantum gravity (see e.g. \cite{Hod2}, and references therein). 
Furthermore, the anti-de Sitter/conformal field theory (AdS/CFT) 
conjecture \cite{Mal} has led to an intensive study of 
black-hole QNM in asymptotically AdS spacetimes in the last few years 
\cite{Hor,Chan,Car1,Car2,Bir,Kon,Star,Son,Aros,Moss,Mus,BerKokA}. 
According to the AdS/CFT correspondence, a large static black hole in AdS corresponds to 
an approximately thermal state in CFT. Perturbing the black hole corresponds to perturbing this 
thermal state, and the decay of the perturbation (characterized by QNM ringdown) 
describes the return to thermal equilibrium.

The dynamics of black-hole perturbations is governed by the Regge-Wheeler equation \cite{RegWheel} 
in the case of a Schwarzschild black hole, and by the Teukolsky equation \cite{Teukolsky} for the Kerr black hole. 
The black hole QNMs correspond to solutions of the wave equations with the physical boundary
conditions of purely outgoing waves at spatial infinity 
and purely ingoing waves crossing the event horizon \cite{Detwe}. Such boundary 
conditions single out a {\it discrete} set of black-hole 
resonances $\{\omega_n\}$ (assuming a time dependence of the 
form $e^{-i\omega t}$). 

Since the perturbation field can fall into the black hole or radiate to infinity, the perturbation decays 
and the corresponding QNM frequencies are {\it complex}. 
It turns out that there exist an infinite number of quasinormal modes, 
characterizing oscillations with decreasing relaxation times (increasing imaginary part) \cite{Leaver,Bach}. 
The mode with the smallest imaginary part (known as the {\it fundamental} mode) gives the 
dynamical timescale $\tau$ for generic perturbations to decay 
(the relaxation time required for the perturbed black hole to return to a quiescent state). 
Namely, $\tau \equiv \omega^{-1}_I$, where $\omega_I$ denotes 
the imaginary part of the fundamental, least damped black-hole resonance. 

Taking cognizance of the relaxation bound Eq. (\ref{Eq3}), 
one finds an upper bound on the black-hole fundamental quasinormal frequency,

\begin{equation}\label{Eq4}
\omega_I \leq \pi T_{BH}/\hbar \  .
\end{equation}
This relation implies that every black hole should have (at least) one QNM resonance 
whose imaginary part conform to the upper bound Eq. (\ref{Eq4}), 
and which determines the characteristic relaxation timescale of the perturbed black hole \cite{Notedim}.

\begin{figure}[tbh]
\centerline{\epsfxsize=9cm \epsfbox{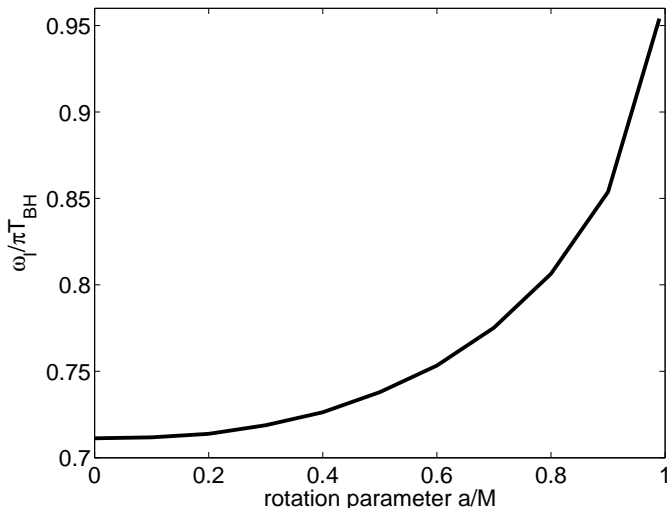}} 
\caption{Imaginary part of Kerr black-hole QNM frequencies as a function of the black-hole 
rotation parameter $a$. The numerical results are for the fundamental (least damped) 
gravitational resonances with $l=m=2$. 
The resonant frequencies conform to the bound $\omega_{I} \leq \pi T_{BH}/\hbar$, 
and saturate it in the extremal limit $a/M \to 1$.}
\label{Fig1}
\end{figure}

\begin{figure}[tbh]
\centerline{\epsfxsize=9cm \epsfbox{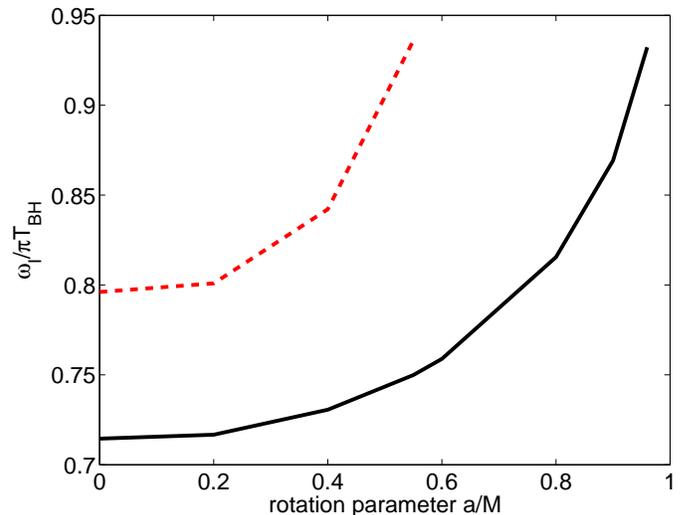}} 
\caption{Imaginary part of Kerr-Newman black-hole QNM frequencies as a function of the black-hole 
rotation parameter $a$. We display results for $Q/M=0.2$ (lower curve) and $Q/M=0.8$ (upper curve, for 
which $a/M \leq 0.6$). 
The numerical results are for the fundamental resonances with $l=m=2$. 
The resonant frequencies conform to the bound $\omega_{I} \leq \pi T_{BH}/\hbar$, 
and saturate it in the extremal limit $a^2+Q^2 \to M^2$.}
\label{Fig2}
\end{figure}

{\it Analytical and numerical confirmation of the bound.---} 
We now confirm the validity of the universal relaxation bound, starting 
with the `canonical' Kerr black holes. 
Figure \ref{Fig1} displays the quantity $\hbar \omega_I/\pi T_{BH}$ for the least damped (fundamental) 
quasinormal frequencies of rotating Kerr black holes \cite{Leaver,Det,Ono,GlaAnd}. 
In contrast with laboratory sized systems, black holes are found to have 
relaxation times which are of the same order of magnitude as the minimally allowed 
one, $\tau_{min}$. Specifically, one finds (see Fig. \ref{Fig1}) that the fundamental black-hole 
resonances are characterized by $\hbar \omega_{I}/\pi T_{BH} \lesssim 1$ \cite{Note2}. 
In Fig. \ref{Fig2} we depict similar results for Kerr-Newman black holes \cite{BerKok}. 
Remarkably, extremal black holes (which are characterized by $T_{BH} \to 0$) {\it saturate} 
the dynamical relaxation bound. Namely, their relaxation times are infinitely long, in accord with the 
third-law of thermodynamics.

It is of great interest to check the validity of the upper bound, Eq. (\ref{Eq4}), 
for other black-hole spacetimes. 
The endpoint of a charged and non-rotating gravitational collapse is described by the 
Reissner-Nordstr\"om spacetime. 
The black-hole formation is followed by a relaxation phase, which describes the 
decay of two types of perturbation fields: coupled gravitational-electromagnetic perturbations, 
and charged perturbation fields. One finds \cite{HodIn} that the fundamental QNM frequency of 
a charged scalar field is characterized by 
$\omega_I = \pi T_{BH}$ in the $eQ/T_{BH} \gg 1$ limit, where $e$ is the charge coupling constant. 
Thus, a charged Reissner-Nordstr\"om black hole saturates the relaxation bound.

The QNM of a D-dimensional Schwarzschild black hole can be calculated analytically in the limit of 
a large angular index $\ell \gg 1$ \cite{KonD}. Using the results of \cite{KonD}, we find

\begin{equation}\label{Eq5}
{{\hbar \omega_I} \over {\pi T_{BH}}}=\Big({2 \over {D-1}}\Big)^{1/(D-3)} {2 \over {\sqrt{D-1}}}\  ,
\end{equation}
which is a monotonic decreasing function of $D$, with a maximum 
of $\hbar \omega_I/\pi T_{BH}=4/(3\sqrt{3})<1$ at D=4. 
Thus, D-dimensional Schwarzschild black holes conform to the upper bound Eq. (\ref{Eq4}).

Furthermore, D-dimensional Schwarzschild-de Sitter black holes \cite{Van} 
also conform to the upper bound Eq. (\ref{Eq4}), with the desired property of 
saturating it in the extremal limit, $T_{BH} \to 0$. 
Large black holes in asymptotically AdS spacetime (large compared to the AdS radius) 
have the interesting property that for odd parity gravitational perturbations, 
the imaginary parts of the QNM frequencies scale like the inverse of the 
black-hole temperature (in this regime, $T_{BH} \sim r_+$) \cite{Car2,BerKokA}. 
Hence, these modes are particularly long lived 
and conform to the relaxation bound Eq. (\ref{Eq4}). 
In the small AdS black-hole regime, the imaginary parts of the quasinormal 
frequencies scale with $r^2_+$ \cite{Car2}. Thus, these black-hole resonances 
also conform to the relaxation bound.

{\it Summary.---} 
In this Letter we have derived a universal bound on relaxation times of 
perturbed thermodynamic systems, $\tau \geq \hbar/\pi T$ \cite{NoteNam}. 
The relaxation bound is a direct consequence of quantum information theory and 
thermodynamic considerations.

We conjecture that a relation of this form could serve as a quantitative way to express the 
third-law of thermodynamics. Namely, one cannot reach a temperature $T$ in a timescale shorter 
than $\hbar/\pi T$ (which indeed goes to infinity in the limit of absolute zero of 
temperature, in accord with the third-law).

Remarkably, black holes comply with the dynamical bound; in fact they have 
relaxation times which are of the same order of magnitude as $\tau_{min}$, 
the minimal relaxation time allowed by quantum theory. 
Moreover, extremal black holes (in the $T_{BH} \to 0$ limit) actually attain 
the bound-- their relaxation time is infinitely long. 
Since black holes {\it saturate} the fundamental bound, we conclude that 
when judged by their relaxation properties, black holes are the most extreme objects in nature.

\bigskip
\noindent
{\bf ACKNOWLEDGMENTS}
\bigskip

I thank Uri Keshet for numerous discussions, as well as for 
a continuing stimulating collaboration. 
I also thank Oren Shriki and Oded Hod for helpful discussions.


\begin{thebibliography}{99}
\bibitem{Noteln2} There is a factor of $\ln 2$ to convert entropy natural units to bits.
\end{thebibliography}
\end{document}